\documentstyle[11pt,aaspp4]{article}

\slugcomment{To appear in {\it Astrophys. J.}}

\begin{document}

\title{Experimentally Derived Dielectronic Recombination Rate
Coefficients for Heliumlike C~V and Hydrogenic O~VIII}

\author{D. W. Savin}
\affil{Department of Physics and Columbia
Astrophysics Laboratory, Columbia University, \\ New York, NY 10027, USA}
\authoremail{savin@astro.columbia.edu}

\begin{abstract}

Using published measurements of dielectronic recombination (DR)
resonance strengths and energies for C~V to C~IV and O~VIII to O~VII,
we have calculated the DR rate coefficient for these ions.  Our derived
rates are in good agreement with multiconfiguration,
intermediate-coupling and multiconfiguration, fully-relativistic
calculations as well as with most $LS$ coupling calculations.  Our
results are not in agreement with the recommended DR rates commonly
used for modeling cosmic plasmas.  We have used theoretical radiative
recombination (RR) rates in conjunction with our derived DR rates to
produce a total recombination rate for comparison with unified RR+DR
calculations in $LS$ coupling.  Our results are not in agreement with
undamped, unified calculations for C V but are in reasonable agreement
with damped, unified calculations for O VIII.  For C~V, the Burgess
general formula (GF) yields a rate which is in very poor agreement with
our derived rate.  The Burgess \& Tworkowski modification of the GF
yields a rate which is also in poor agreement.  The Merts et
al.\ modification of the GF yields a rate which is in fair agreement.
For O~VIII the GF yields a rate which is in fair agreement with our
derived rate.  The Burgess \& Tworkowski modification of the GF yields
a rate which is in good agreement.  And the Merts et al.\ modification
yields a rate which is in very poor agreement.  These results suggest
that for $\Delta N=1$ DR it is not possible to know {\it a priori}
which formula will yield a rate closer to the true DR rate.  We
describe the technique used to obtain DR rate coefficients from
laboratory measurements of DR resonance strengths and energies.  
For use in plasma modeling, we also present easy-to-use fitting formulae 
for the experimentally derived DR rates.

\end{abstract}


\section{Introduction}
\label{sec:Introdution}

Carbon and oxygen are two of the most abundant elements in the universe
and lines from these elements provide valuable plasma diagnostics for
almost all classes of cosmic sources.  Essential for many of these
diagnostics are accurate electron-ion recombination rate coefficients,
particularly of dielectronic recombination (DR), which for most ions in
electron-ionized plasmas is the dominant means of electron capture
(Arnaud \& Rothenflug 1985).  Producing accurate theoretical DR rate
coefficients is, however, theoretically and computationally
challenging.

In the past, semi-empirical expressions such as the Burgess (1965)
formula along with modified versions by Burgess \& Tworkowski (1976)
and Merts et al.\ (1976) were developed to calculate DR rates.  More
recently, a number of more sophisticated theoretical approaches have
been used to calculate DR, among them single-configuration
$LS$-coupling (Bellantone \& Hahn 1989), multiconfiguration
intermediate-coupling (Pindzola, Badnell, \& Griffin 1990), and
multiconfiguration fully-relativistic (Chen 1988) techniques, as well
as undamped and damped, unified radiative recombination (RR) and DR
calculations in $LS$-coupling (Nahar \& Pradhan 1997; Nahar 1999).
Approximations, though, need to be made to make any of these techniques
computationally tractable (Hahn 1993).  Currently, sophisticated DR
calculations are non-existent for many ions, and in the absence of
anything better, semi-empirical formulae are often still used for
plasma modeling.

Laboratory measurements can be used to test the different theoretical
and computational techniques for calculating DR.  Recently, Savin et
al.\ (1997, 1999) developed a technique for obtaining rate coefficients
from laboratory measurements of DR resonance strengths and energies.
They successfully used this technique to derive rates for $\Delta n=0$
DR of Fe XVIII and Fe XIX and to benchmark existing theoretical
calculations.  Here, we describe this technique in detail for the first
time and apply it to recent DR measurements in C~V and O~VIII.
Kilgus et al.\ (1990, 1993) and Mannervik et al.\ (1997) have measured
the resonance strengths and energies for DR of C~V to C~IV and O~VIII
to O~VII.  We use their results to produce DR rate coefficients to
benchmark existing C~V and O~VIII DR calculations and to provide rates
for use in plasma modeling.  

In electron-ionized plasmas, lines from heliumlike C~V and hydrogenic
O~VIII trace gas at $T_e \sim 10^{5.1-5.9}$ K and $\sim 10^{6.4}$ K,
respectively (Arnaud \& Rothenflug 1985; Mazzotta et al.\ 1998).  C~V
and O~VIII lines have been observed in solar spectra (Doschek \& Cowan
1984) and O~VIII lines in supernova remnants (Winkler et al.\ 1981).
And with the upcoming launches of {\it Chandra} and {\it XMM} and the
high-resolution spectrometers aboard, C~V and O~VIII lines are expected
to be seen in may other electron-ionized, cosmic sources.

Using different heavy-ion storage rings, Kilgus et al.\ (1993) and
Mannervik et al.\ (1997) have measured DR for C~V via the capture
channels
\begin{equation}
{\rm C}^{4+}(1s^2) + e^- \rightarrow {\rm C}^{3+}(1s2lnl^\prime)
\ (n=2,\ldots,n_{max}).
\end{equation}
where $n_{max} \sim 28$ for the results of Kilgus et al. and $\sim 16$
for the results of Mannervik et al.  Kilgus et al.\ (1990) have also
measured DR for O~VIII via the capture channels
\begin{equation}
{\rm O}^{7+}(1s) + e^- \rightarrow {\rm O}^{6+}(2lnl^\prime)
\ (n=2,\ldots,n_{max})
\end{equation}
where $n_{max}\sim 69$.  The radiative stabilization of these
autoionizing C~V and O~VII states to bound configurations results in
DR.  Details of the experimental techniques used are given in the
references cited.

The paper is organized as follows:  We describe in Section
\ref{sec:MethodofCalculation} how one produces a DR rate coefficient
using measured DR resonance strengths and energies.  In Section
\ref{sec:ResultsandDiscussion} we present the resulting rate
coefficients and compare the derived DR rates with published theoretical
rates.  We also give a simple fitting formula for use in plasma
modeling.

\section{Method of Calculation} 
\label{sec:MethodofCalculation}

DR is a resonance process consisting, in the zero-density limit,
of an infinite number of
resonances.  The DR rate coefficient $\alpha$ for a plasma with a
Maxwell-Boltzmann electron distribution is given by
\begin{equation}
\alpha(T_e)=\int \sum_d \sigma_d(E) v_e(E) P(T_e,E) dE
\label{eq:rate1}
\end{equation}
where $T_e$ is the electron temperature; $\sigma_d(E)$ is the
energy-dependent DR cross section for resonance $d$; $v_e(E)$ is the
relative electron-ion velocity at energy $E$, which is taken to be the
electron energy as the ions are nearly to stationary in the
center-of-mass frame; and the sum is over all DR resonances.  The
Maxwell-Boltzmann distribution $P(T_e,E)$ is given by
\begin{equation}
P(E,T_e)dE ={2 E^{1/2} \over \pi^{1/2} (k_BT_e)^{3/2}} \exp\Biggl( {-E\over
k_BT_e}\Biggr)dE
\label{eq:pofe}
\end{equation}
where $k_B$ is the Boltzmann constant.

Kilgus et al.\ (1990, 1993) and Mannervik et al.\ (1997) published
measured DR resonance strengths $\hat{\sigma}_d$ and energies
$E_d$.  The DR resonance strength is defined
\begin{equation}
\hat{\sigma}_d(E)=
\int_{E-\delta E/2}^{E+\delta E/2} \sigma_d(E^\prime) dE^\prime
\label{eq:strength1}
\end{equation}
where $\sigma_d$ is the cross section for a resonance or group of
resonances labeled $d$ and \{$E-\delta E$,$E+\delta E$\} is a region in
energy chosen such that it contains only those resonances comprising
$d$.
 
Here we are interested in calculating rate coefficients.  This involves
convolving the DR resonances with the slowly varying function
$P(T_e,E)$.  Because the energy widths of the measured resonances are
smaller than the scale over which $P(T_e,E)$ changes, for our purposes
we can accurately approximate $\sigma_d(E)$ as
\begin{equation}
\sigma_d(E)=\hat{\sigma}_d(E_d)\delta(E-E_d)
\label{eq:cross1}
\end{equation}
where $E_d$ is the energy of resonance $d$ and $\delta(E-E_d)$ is the Dirac
delta function.  The DR rate coefficient for Maxwellian plasmas is
found by substituting Equation \ref{eq:cross1} into Equation
\ref{eq:rate1} which yields
\begin{equation}
\alpha(T_e)=\sum_d \hat{\sigma_d}(E_d) v_e(E_d) P(T_e,E_d).
\label{eq:rate2}
\end{equation}

Kilgus et al.\ (1993) and Mannervik et al.\ (1997) do not report
measured resonance energies for capture by C~V into levels where $n \ge 4$.
To calculate these resonance energies $E_n$ we use the Rydberg formula
\begin{equation}
E_n=\Delta E - q^2R_\infty/n^2
\label{eq:rescond}
\end{equation}
where $q=4$ is the charge of the ion before recombination,
$\Delta E=307.8$ eV is the energy of the $1s^2 (^1S_0) - 1s2p
(^1P_1)$ core excitation (Kelly 1987), and $R_\infty$ is the Rydberg
constant.  For O~VIII, $q=7$ and $\Delta E=653.6$ eV is the
energy of the $1s (^1S_{1/2}) - 2p (^1P_{3/2,1/2})$ core excitation
(Kelly 1987).

Mannervik et al.\ (1997) estimate that they measured DR for capture into
levels $n_{max}\lesssim16$.  Kilgus et al.\ (1993) do not report a value of
$n_{max}$, so we derive an estimate here.  Using their ion energy, the
bending radius of the dipole magnets in their experiment (Linkemann
1995), and the semiclassical formula for field ionization (Brouillard
1983), we estimate electrons captured into levels where $n = n_{cut}
\gtrsim 19$ will be field ionized and thus not detected.  However, the
captured electrons can radiatively decay below $n_{cut}$ during the
$\sim 5.1$ m distance they travel between the electron-ion interaction
region in the experiment and the dipole magnet (cf., Kilgus et al.\ 1992;
Savin et
al.\ 1997).  Using the ion velocity, the hydrogenic formula for
radiative lifetimes of Marxer \& Spruch (1991), and calculations which
show that DR for heliumlike ions essentially populates only angular
momentum levels where $l\le4$ (Chen 1986), we estimate that electrons
captured into levels $n_{max} \lesssim 28$ will radiatively decay below
$n_{cut}$ and thus were detected by Kilgus et al.  For OVIII, Kilgus et
al.\ (1990) estimate $n_{max}\sim 69$.

For a given DR series, as $n$ increases the energy spacing between DR
resonances decreases.  Due to the energy resolutions of the experiments, 
above some $n$ level it is not possible to resolve the DR resonances and
determine individual values of $\sigma_n$.  For these high $n$ levels,
Kilgus et al.\ (1990; 1993) presents total resonance strengths summed
from $n\ge9$ to $n_{max}$ for C~V and from $n\ge8$ to $n_{max}$ for
O~VIII.  We divide and spread out these summed resonance strengths into
$\sim 1$ eV wide bins when using Equation \ref{eq:rate2}.  The
``resonance'' energies of these bins are chosen to lie between $E_9$
and $E_{n_{max}}$ for C~V and between $E_8$ and $E_{n_{max}}$ for
O~VIII.  The bin widths are smaller than the scale over which $P(T_e,E)$
changes and the small errors in the exact resonance energies of these high
$n$ levels has an insignificant affect on the derived DR rates.

\section{Results and Discussion}
\label{sec:ResultsandDiscussion}

\begin{figure}
\plotone{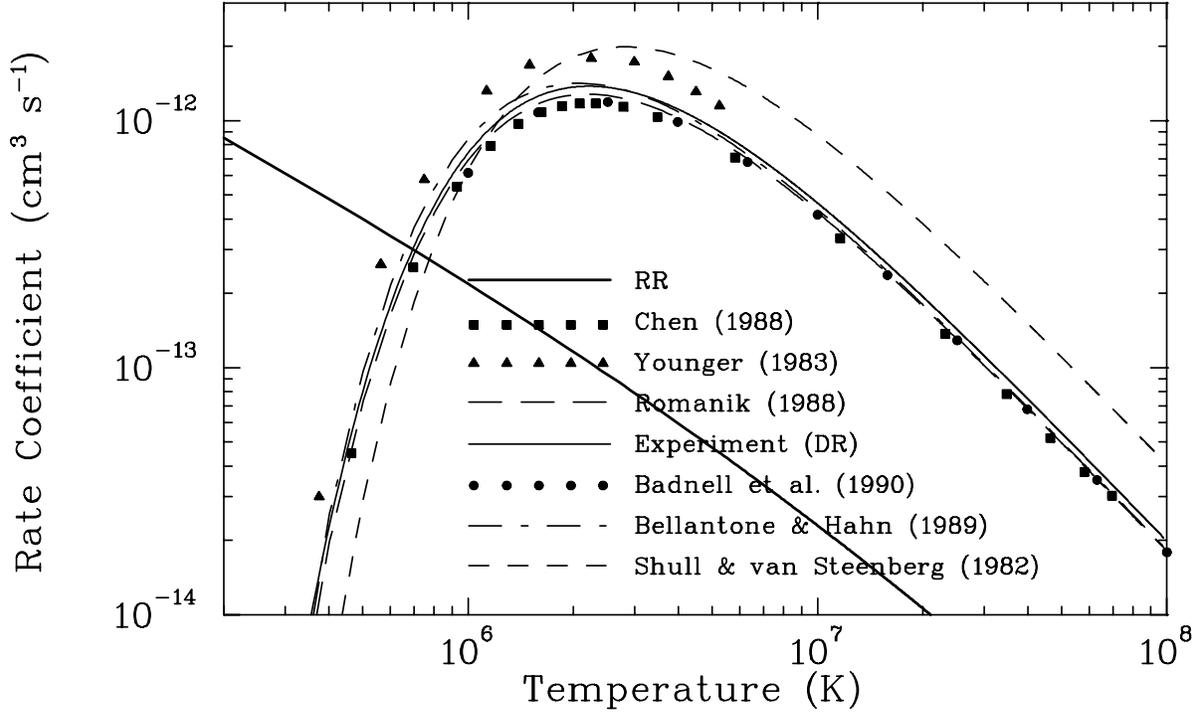}
\caption{C~V to C~IV Maxwellian-averaged DR rate coefficients.
The thin solid curve is the integration of the experimental DR
resonance strengths and energies from Kilgus et al.\ (1993) and
Mannervik et al.\ (1997).  The short-dashed curve the recommended rate
of Shull \& van Steenberg (1982).  The long-dashed curve is the
single-configuration, $LS$-coupling results of Romanik (1988) and the
long-dashed-dot curve of Bellantone \& Hahn (1989).  The triangles are
the $LS$-coupling calculations of Younger (1983); the circles the
intermediate-coupling, multiconfiguration calculations of Badnell et
al.\ (1990); and the squares the fully-relativistic, multiconfiguration
calculations of Chen (1988).  The thick
solid line is the RR rate of Verner \& Ferland (1996).}
\label{fig:CV}
\end{figure}

\begin{figure}
\plotone{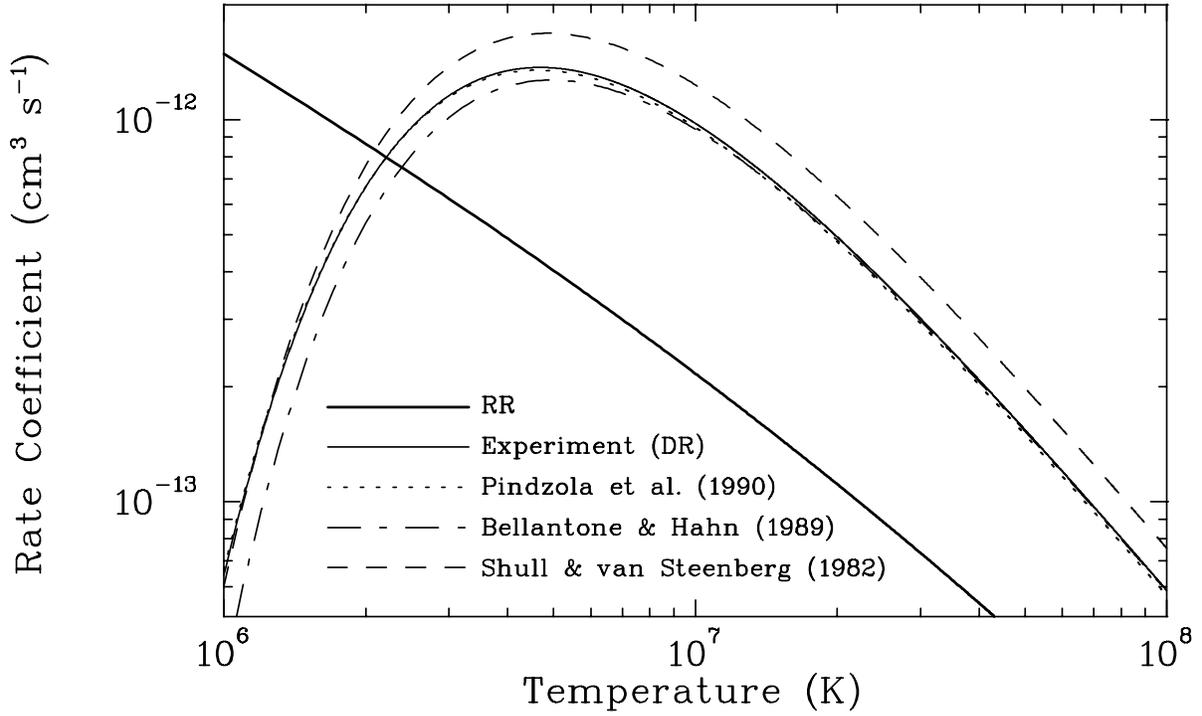}
\caption{O~VIII to O~VII Maxwellian-averaged DR rate
coefficients.  The thin solid curve is the integration of the
experimental DR resonance strengths and energies from Kilgus et
al.\ (1990).  The short-dashed curve the recommended rate of Shull \&
van Steenberg (1982).  The long-dashed-dot curve is the
single-configuration, $LS$-coupling results of Bellantone \& Hahn
(1989) and the dotted curve the intermediate-coupling,
multiconfiguration calculations of Pindzola et al. (1990).  The thick
solid line is the RR rate of Verner \& Ferland (1996).}
\label{fig:OVIII}
\end{figure}

In Figures \ref{fig:CV} and \ref{fig:OVIII} we present the C~V and O
VIII DR rate coefficients, respectively, derived using the measured
resonance strengths and energies of Kilgus et al.\ (1990, 1993) and
Mannervik (1997).  The unmeasured contribution to the DR rate due to capture
into levels where $n \gtrsim n_{max}$ is predicted to be insignificant
(Chen 1986; Pindzola et al.\ 1990).  The absolute uncertainties in the
derived DR rates are estimated to be $\lesssim \pm25\%$ which corresponds
to the reported absolute experimental uncertainties.

Existing theoretical DR rates are also shown in Figures \ref{fig:CV}
and \ref{fig:OVIII} as well as the RR rates of Verner \& Ferland
(1996).  For C~V, the single-configuration $LS$-coupling calculations
of Bellantone \& Hahn (1986) and Romanik (1988), multiconfiguration
intermediate-coupling calculations of Badnell, Pindzola, \& Griffin
(1990), and multiconfiguration fully-relativistic calculations of Chen
(1988) are all in good agreement with our experimentally inferred rate.
The $LS$-coupling calculations of Younger (1983) are $\sim 30\%$ larger
than the experimental rate.  The recommended and commonly used DR rate
of Shull \& van Steenberg (1982) peaks at a value $\sim 45\%$ larger
than ours and has a steeper low $T_e$ behavior.

Pindzola et al.\ (1990) calculated resonance strengths for O VIII using
intermediate-coupling, multiconfiguration Hartree-Fock (MCHF) and
intermediate-coupling, multiconfiguration Thomas-Fermi (MCTF)
techniques.  Using their MCHF results for capture into levels $n\le6$
and their MCTF results for $n\ge7$, as well as the experimental
resonance energies, we calculate the corresponding DR rate using
Equation \ref{eq:rate2}.  The resulting rate is in good agreement with
our derived rate.  The single-configuration $LS$-coupling rate of
Bellantone \& Hahn (1986) is also in good agreement with the
experimental rate, though with decreasing $T_e$ their rate falls off
sooner than ours.  The recommended and commonly used DR rate of Shull
\& van Steenberg (1982) is $\sim 31\%$ larger than our derived rate.
Not shown in Figure \ref{fig:OVIII}, for reasons of clarity, is the
rate of Zhdanov (1978) whose peak rate is $\sim 7$ times larger than
the experimental rate.

\begin{figure}
\plotone{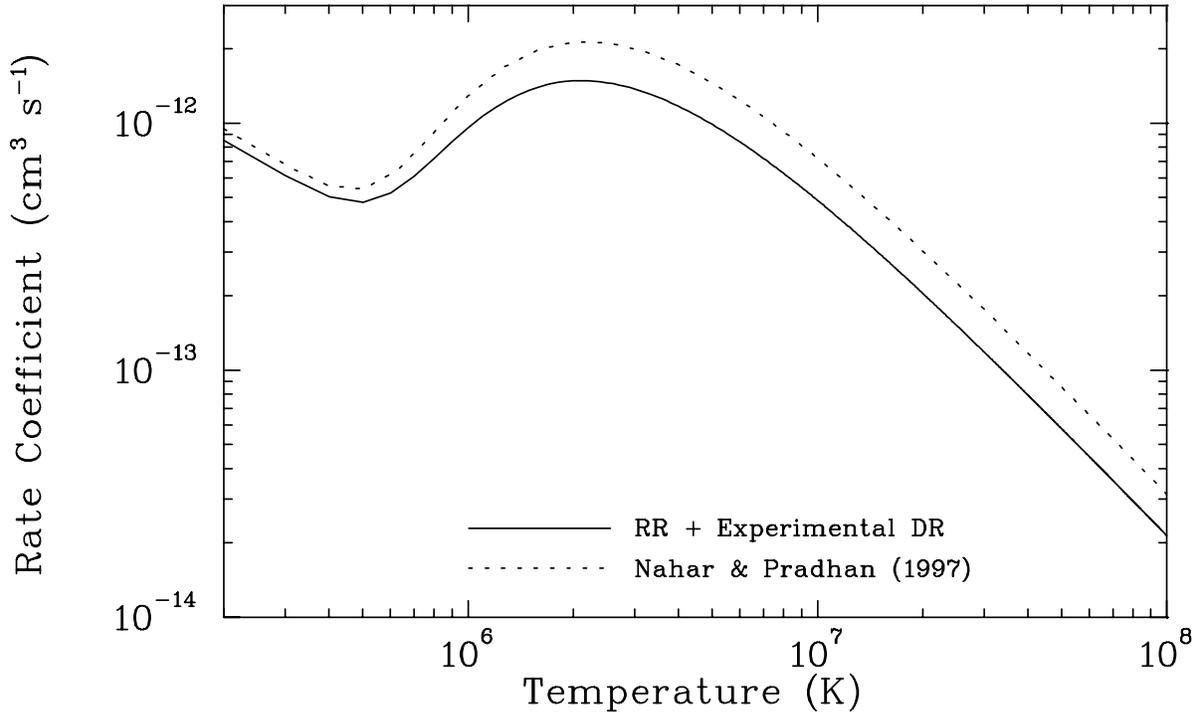}
\caption{C~V to C~IV Maxwellian-averaged RR+DR rate coefficients.  The
solid curve is sum of the RR rate from Verner \& Ferland (1996)
plus the DR rate derived from the integration of the experimental DR
resonance strengths and energies from Kilgus et al.\ (1993) and
Mannervik et al.\ (1997).  The dotted curve is the undamped, unified RR+DR
calculations in $LS$-coupling of Nahar \& Pradhan (1997).}
\label{fig:CVNah}
\end{figure}

\begin{figure}
\plotone{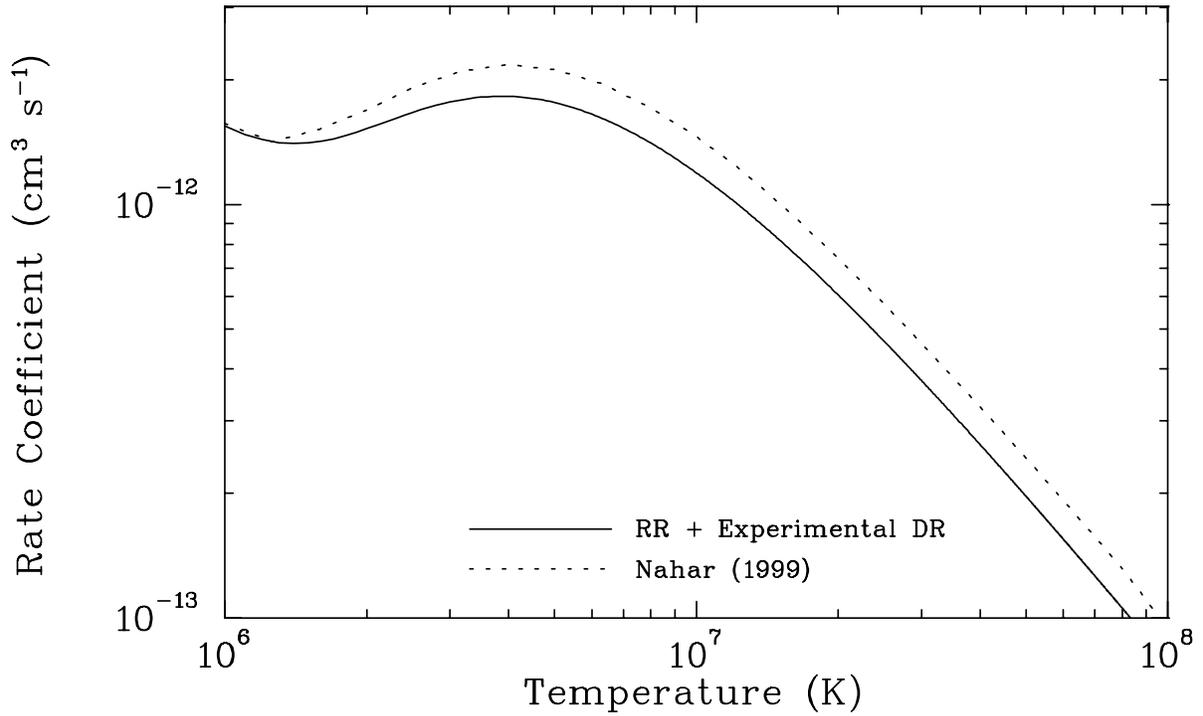}
\caption{O~VIII to O~VII Maxwellian-averaged RR+DR rate coefficients.  The
solid curve is sum of the RR rate from Verner \& Ferland (1996)
plus the DR rate derived from the integration of the experimental DR
resonance strengths and energies from Kilgus et al.\ (1990).  
The dotted curve is the damped, unified RR+DR
calculations in $LS$-coupling of Nahar (1999).}
\label{fig:OVIIINah}
\end{figure}

Nahar \& Pradhan (1997) and Nahar (1999) present unified RR+DR rates.
To compare with their results we add the RR rates of Verner \& Ferland
(1996) to the experimentally derived DR rates.  Thus we treat RR and DR
as independent processes and do not allow for the possibility of
interference between the two recombination channels.  The validity of
this approach is supported by recent theoretical work (for DR on systems
ranging in complexity from C II and Mg II to U XCII) which has shown the
effect of interferences to be small (\cite{Pind92a}).
Figures~\ref{fig:CVNah} and \ref{fig:OVIIINah} show for C V and O VIII,
respectively, the unified RR+DR rate of Nahar \& Pradhan and the total
RR+DR rate using the derived DR results and the RR results of Verner \&
Ferland.  For C V, at peak value the undamped, unified rate of Nahar \&
Pradhan is $\sim 43\%$ larger than our resulting total RR+DR rate.
This is consistent with the estimate by Pradhan \& Zhang (1997) that
allowance for radiation damping would reduce the undamped rate by $\sim
20-30\%$.  For O VIII, at peak value the damped, unified rate of Nahar
is $\sim 20\%$ larger than our resulting total RR+DR rate.

\begin{figure}
\plotone{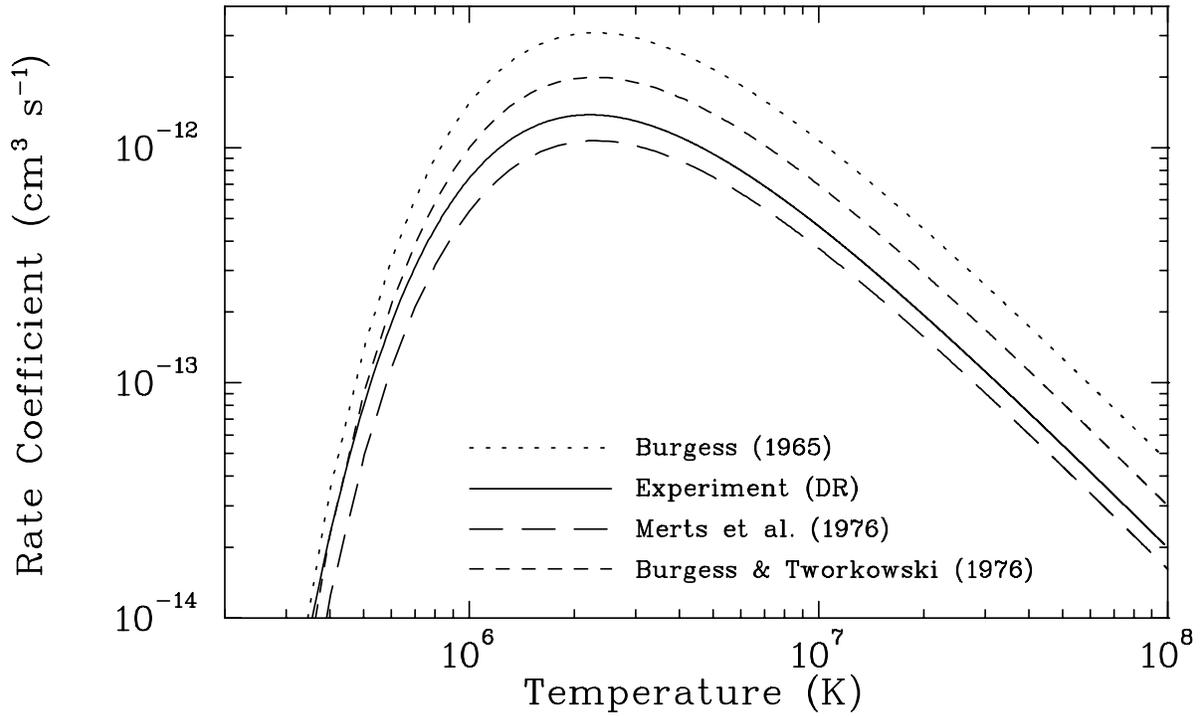}
\caption{C~V to C~IV Maxwellian-averaged DR total rate coefficients.
The solid curve is the integration of the experimental DR
resonance strengths and energies from Kilgus et al.\ (1993) and
Mannervik et al.\ (1997).  The dotted curve is the Burgess (1965)
formula, the long-dashed curve the Merts et al.\ (1976) formula, and
the short-dashed curve the Burgess \& Tworkowski (1976) formula.}
\label{fig:CVBurg}
\end{figure}

\begin{figure}
\plotone{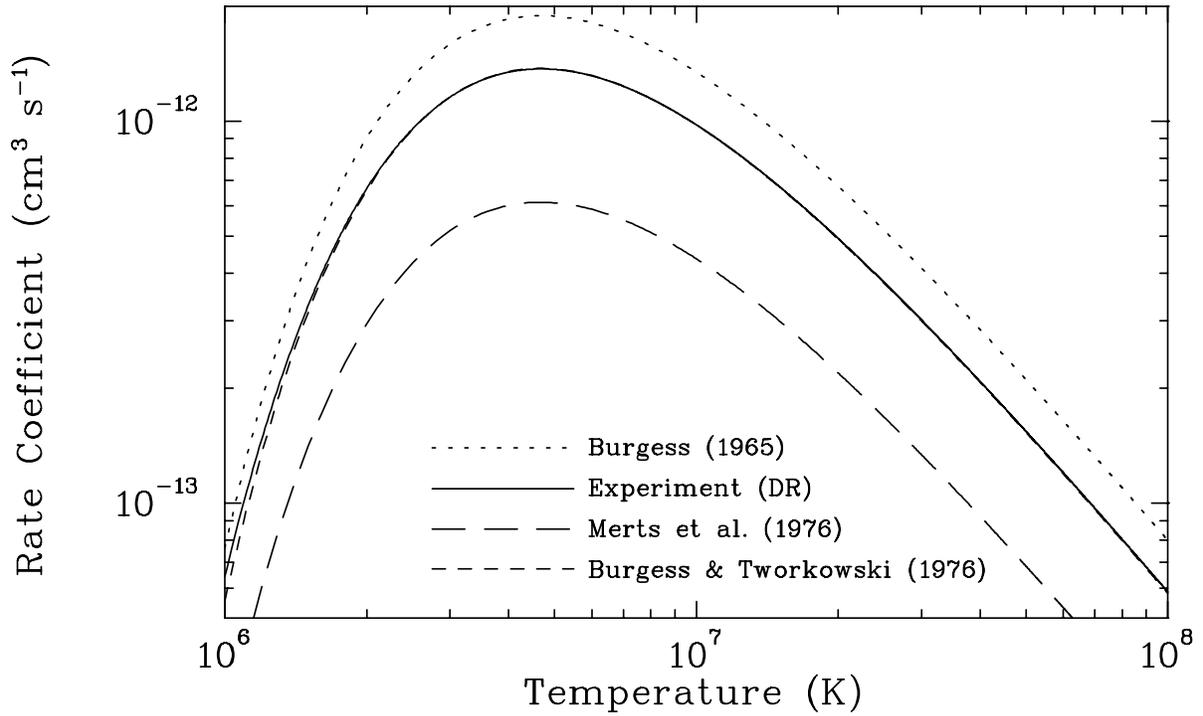}
\caption{O~VIII to O~VII Maxwellian-averaged DR total rate
coefficients.  The solid curve is the integration of the
experimental DR resonance strengths and energies from Kilgus et
al.\ (1990).  The dotted curve is the Burgess (1965)
formula, the long-dashed curve the Merts et al.\ (1976) formula, and
the short-dashed curve the Burgess \& Tworkowski (1976) formula.}
\label{fig:OVIIIBurg}
\end{figure}

DR rate coefficients are sometimes calculated using the Burgess (1965)
general formula (GF) or versions of the GF as modified by Merts et
al.\ (1976) and by Burgess \& Tworkowski (1976).  We have calculated
the Burgess and Merts et al.\ rates using the formulae as given by
Cowan (1981).  We use oscillator strengths and excitation energies from
Wiese, Smith, \& Glennon (1966).  Shown in Figure \ref{fig:CVBurg} for
C~V is the Merts et al.\ rate which is in fair agreement with our
derived rate, though with decreasing $T_e$ it goes to zero sooner than
the experimental rate.  Also shown in Figure \ref{fig:CVBurg} is the
Burgess GF rate which is a factor of $\sim 2.2$ times larger than the
experimental rate.  The Burgess \& Tworkowski (1976) modification of
the GF yields a rate $\sim 50\%$ larger than our derived rate.  This is
surprisingly good considering that their modification is meant for DR
forming heliumlike, not lithiumlike, ions.  Figure \ref{fig:OVIIIBurg}
shows for O VIII the GF rate which is $\sim 38\%$ larger than the
experimental rate and, considering the expected accuracy of the Burgess
formula, in fair agreement.  The Burgess \& Tworkowski rate is in good
agreement.  Also shown is the Merts et al.\ formula rate which is a
factor of $\sim 2.2$ smaller than the experimental rate.  Our C V and O
VIII results strongly suggest that for $\Delta N=1$ DR it is not
possible {\it a priori} to know which formula will yield a result
closer to the true rate.

For use in plasma modeling, we have fit the experimentally derived C~V
and O~VIII DR rates using the simple formula (Arnaud \& Raymond 1992)
\begin{equation} 
\alpha(T_e)=T_e^{-3/2}\sum_i c_i e^{-E_i/k_BT_e}.  
\label{eq:drratefit} 
\end{equation} 
Here $c_i$ and $E_i$ are, respectively, the strength and energy
parameters for the $i$th fitting component.  Best fit values are listed
in Table \ref{tab:fitparameters}.  For C~V, the fit is good to better
1\% for $2.6\times10^5 \le T_e \le 10^8$ K.  Below $2.6\times10^5$ K,
with decreasing $T_e$ the fit goes to zero faster than the derived
rate.  But this error is insignificant for plasma modeling as the RR
rate is $\sim 3$ orders of magnitude larger than the DR rate at these
temperatures.  For O~VIII, the fit is good to better 1\% for
$7.1\times10^5 \le T_e \le 10^8$ K.  Below $7.1\times10^5$ K, with
decreasing $T_e$ the fit goes to zero faster than the derived rate.
This error is insignificant for plasma modeling as the RR rate is
$\sim 2$ orders of magnitude larger than the DR rate at these
temperatures.

\section{Summary}
\label{sec:Summary}

We have presented a simple technique for obtaining DR rate coefficients
from laboratory measurements of DR resonance strengths and energies.
With this technique, we have derived DR rates for C~V to C~IV and
O~VIII to O~VII using published resonance strengths and energies.  Our
derived rates are in good agreement with multiconfiguration,
intermediate-coupling and multiconfiguration, fully-relativistic
calculations as well as with most $LS$-coupling calculations.  Our
rates are not in agreement with the recommended DR rates commonly used
for astrophysical plasma modeling.  We have used theoretical radiative
recombination (RR) rates in conjunction with our derived DR rates to
produce a total recombination rate for comparison with unified RR+DR
calculations in $LS$ coupling.  Our results are not in agreement with
undamped, unified calculations for C V but are in reasonable agreement
with damped, unified calculations for O VIII.  Also, neither the
Burgess general formula, the Merts et al.\ formula, nor the Burgess \&
Tworkowski formula consistently yield a rate which is in agreement with
our derived rates.  This suggests that for $\Delta n=1$ DR it is not
possible to know {\it a priori} which formula will yield a rate closer
to the true DR rate.  We have also presented simple fitting formula of
the experimentally derived DR rates for use in plasma modeling.

\acknowledgements

The author would like to thank N. R. Badnell, M. H. Chen, T. W.
Gorczyca, S. M. Kahn, D.  A.  Liedahl, S. N. Nahar, and A. Wolf for
stimulating conversation and for critically reading the manuscript.
This work was supported in part by NASA High Energy Astrophysics X-Ray
Astronomy Research and Analysis grant NAG5-5123.

\vfill
\clearpage
\eject

\begin{deluxetable}{cccccc}
\footnotesize
\tablecaption{Fit parameters for the experimentally
derived C~V to C~IV and O~VIII to O~VII DR rate coefficient.
The units for $c_i$ are cm$^3$~s$^{-1}$~K$^{1.5}$ and for
$E_i$ are eV.
\label{tab:fitparameters}}
\tablewidth{0pt}
\tablehead{
\colhead{ } & \multicolumn{2}{c}{C V} & \colhead{ } &
\multicolumn{2}{c}{O VIII} \\
\cline{2-3} \cline{5-6} \\
\colhead{$i$} & \colhead{$c_i$} & \colhead{$E_i$} & & \colhead{$c_i$} 
& \colhead{$E_i$}
}
\startdata
1 & 3.02e-3 & 246 & & 8.56e-3 & 497 \nl
2 & 1.75e-2 & 296 & & 5.43e-2 & 632 \nl
\enddata
\end{deluxetable}

\vfill
\clearpage
\eject

\clearpage
\eject

\end{document}